# Do the propensity and drivers of academics' engagement in research collaboration with industry vary over time?

Giovanni Abramo[1], Francesca Apponi[2] and Ciriaco Andrea D'Angelo[3]

*[1] giovanni.abramo@iasi.cnr.it*
Laboratory for Studies in Research Evaluation, Institute for System Analysis and Computer Science (IASI-CNR), National Research Council of Italy (Italy)

*[2] francesca.apponi@alumni.uniroma2.eu*
Department of Engineering and Management, University of Rome "Tor Vergata" (Italy)

*[3] dangelo@dii.uniroma2.it*
Department of Engineering and Management, University of Rome "Tor Vergata" (Italy)
&
Laboratory for Studies in Research Evaluation, Institute for System Analysis and Computer Science (IASI-CNR), National Research Council of Italy (Italy)

**Abstract**

This study is about public-private research collaboration. In particular, we want to measure how the propensity of academics to collaborate with their colleagues from private firms varies over time and whether the typical profile of such academics change. Furthermore, we investigate the change of the weights of main drivers underlying the academics' propensity to collaborate with industry. In order to achieve such goals, we apply an inferential model on a dataset of professors working in Italian universities in two subsequent periods, 2010-2013 and 2014-2017. Results can be useful for supporting the definition of policies aimed at fostering public-private research collaborations, and should be taken into account when assessing their effectiveness afterwards.

# Introduction

The ability of industry to exploit the results of academic research is a distinctive competence of advanced economies. Policies aimed at developing such ability are among the priorities of an increasing number of governments (Fan, Yang, & Chen, 2015; Shane, 2004). Public-private research collaboration is one of the main modes to realize knowledge transfer. Understanding the motivations underlying joint cooperation is an important step towards formulating policies and initiatives aimed at increasing the frequency of collaboration.

The main objective of this work is to investigate to what extent frequency of public-private research collaboration change over time, alongside the main drivers underlying the academics' propensity to collaborate with industry. Decision makers then would be able to formulate incentive schemes based on those drivers that show not only more weight but also more stability. A critical step in the study is the identification of the main drivers that could influence the propensity of academics to engage in research collaborations with industry. Previous literature suggests that they are to be found in the individual characteristics of the academic and the environment he or she works in (Zhao, Broström, & Cai, 2020; Llopis, Sánchez-Barrioluengo, Olmos-Peñuela, & Castro-Martínez, 2018; Abramo, D'Angelo, & Murgia, 2013), although personal characteristics seem to be more important than those of the environment (D'Este & Patel, 2007).

Most works on the topic are based on surveys, which poses limits on the scale of observations (Zhao, Broström, & Cai, 2020; Weerasinghe & Dedunu, 2020; Llopis, Sánchez-Barrioluengo, Olmos-Peñuela, & Castro-Martínez, 2018; Thune, Reymert, Gulbrandsen, & Aamodt, 2016). To overcome these limits, we adopt instead a bibliometric approach, analyzing the co-authorships of publications. Co-authorship in fact should be the outcome of a real research collaboration, though exceptions cannot be excluded (Katz & Martin, 1997).

We aim to answer the following research questions:
- Does the intensity of public-private research collaboration varies over time?
- Does the typical profile of academics collaborating with the private sector change over time?
- Do the main individual and contextual drivers of the academics' propensity for research collaboration with the private sector change over time?

To address these questions, we apply an inferential model on a dataset of professors working in Italian universities in two subsequent periods, 2010-2013 and 2014-2017. It must be noted that we investigate the scientific output of research collaboration, therefore output at time $t$ refers to a joint research project conducted at time $t$-$\tau$, where $\tau$ represents the time lapse needed for the new knowledge to be encoded in written form and having it published in a scientific journal. This is it be kept in mind in general when interpreting results of time-series or before-after analyses, and especially when setting up frameworks to assess the effectiveness of policies aimed at fostering public-private research collaborations.

In the next section we review the literature on the factors determining private-public research collaboration. In Section three we present the methodology and data. In Section four we show the results of the analysis. Section five concludes the work.

# Literature review

The underlying reasons for academics to engage in research collaborations with industry differ from those of the industrial partner. The latter mainly aims to access knowledge to be further developed for commercial exploitation (Bekkers & Bodas Freitas, 2008; Perkmann et al., 2013). The former is mostly attracted from direct economic and financial benefits that the industrial partner can offer (Garcia, Araújo, Mascarini, Santos, & Costa, 2020) and, to some extent, also by a noble purpose: the so called "mission" motivation, i.e. advancing the societal role of universities (Iorio, Labory, & Rentocchini, 2016). The intensity of public-private research collaboration must be sensitive then to the changing financial needs of academics, to the incentive schemes adopted by national and regional governments and by the research organizations they work at, and to the effectiveness of industry-liaison offices which act as catalysts for collaboration.

From the academic standpoint, the propensity to collaborate with industry varies along the career cycle, although the impact of age seems non-linear (Weerasinghe & Dedunu, 2020). In fact, collaboration propensity seems stronger at the early career stage and weaker in the later stages (Bozeman & Gaughan, 2011; Ubfal & Maffioli, 2011). In fact, younger academics are in more need of financial resources, and to show their ability to activate and manage collaborations, which is positively evaluated for career progress (Bayer & Smart, 1991; Traoré & Landry, 1997).

Gender also plays an important role. Because of gender homophily women show greater difficulty in developing their social capital, including collaboration networks (Boschini & Siogren, 2007). As compared to males, female academics engage less in research collaboration with industry (Tartari & Salter, 2015), mainly because they are less interested (Calvo, Fernández-López, & Rodeiro-Pazos, 2019).

There occurs a positive correlation between research performance and collaboration (Mansfield, 1995; Mansfield & Lee, 1996; He, Geng, & Campbell-Hunt, 2009; Lee & Bozeman, 2005; Schartinger, Schibany, & Gassler, 2001), as companies are likely to search for top players to engage in their research projects (Balconi, & Laboranti, 2006).

The tendency to diversify one's own research activities may underly curiosity for novelty and unexplored, and reveal more appropriate for engaging in cross-sector research (Abramo, D'Angelo, & Di Costa, 2019).

Environmental factors must have an influence as well on the academic's propensity for joint research with industry. The relative importance that the university poses on technology transfer creates a culture and motivational stimuli more or less harbinger of research collaboration with industry (Di Gregorio & Shane, 2003; Giuri, Munari, Scandura, & Toschi, 2019). Also, the extent of financial resources made available for research affect the propensity to search for private sources (Giuri, Munari, Scandura, & Toschi, 2019).

The attitude and inclination of academic colleagues towards cross-sector research projects, will have an influence of one's own behavior, through a mechanism of social comparison (Tartari, Perkmann, & Salter, 2014). The size of the university plays a role in determining the types of collaboration by the research staff. If a critical research mass is not there, academics would likely tend to engage in extra-muros collaborations (Schartinger, Schibany, & Gassler, 2001).

The demand by industry for collaboration, and therefore the concentration of private R&D in the territory of the university, affects the possibilities for cross-sector collaboration (Berbegal-Mirabent, Sánchez García, & Ribeiro-Soriano, 2015).

Finally, the research production function varies across scientific disciplines (Abramo, D'Angelo, & Murgia, 2013a), therefore discipline effects on the intensity of university-industry collaborations are expected.

**Data and methods**

The field of observation consists of professors of Italian universities, conducting research in the so-called hard sciences. We exclude the social sciences and arts & humanities because for these the coverage of bibliographic repertories is still insufficient for reliable representation of research output. In the Italian university system all professors are classified in one and only one field (named the scientific disciplinary sector, SDS, 370 in all). Fields are grouped into disciplines (named university disciplinary areas, UDAs, 14 in all). We observe research production in 201 SDSs falling in 10 UDAs, where publications in international journals serve as a reliable proxy for overall research output.

To investigate whether the collaboration rates and determinants of academic engagement in joint research with industry change over time, we conduct a longitudinal analysis dividing the observed period into two subsequent four-year subperiods, 2010-2013 and 2014-2017.

The analysis dataset consists of all assistant, associate and full professors, on staff for at least three years in 2010-2013 or 2014-2017, with at least one Web of Science (WoS) indexed publication in the relevant period. These are 31634 professors in the first period and 31392 in the second.

The bibliometric dataset is extracted from the Italian Observatory of Public Research, a database developed and maintained by the present authors, and derived under license from the Thomson Reuters WoS. Beginning from the raw data of the WoS and applying a complex algorithm to reconcile the author's affiliation and disambiguation of the true identity of the authors, each publication (article, article review and conference proceeding) is attributed to the professors that authored it (D'Angelo, Giuffrida, & Abramo, 2011).[1] Collaboration with industry is evidenced by the presence of at least one private company[2] in the address list of publications authored by the professor in the dataset.

We will first verify if the share of professors collaborating with industry in the two periods increases or not and if the "profile" of those engaging in such collaborations has changed over time. Then, we will use a logit regression in order to understand if the main drivers of academic engagement in public-private research collaboration changed in weight, in the two periods. The logit regression relies on a dummy dependent variable (Y) assuming: 1, if professor *i* co-authored at least one publication with industry; 0, otherwise. As for covariates suggested by previous literature that are likely to affect the propensity of professors to engage in research collaboration with industry, we will consider the following, grouped in two clusters.

Individual covariates
- Gender ($X_1$), specified by a dummy variable (1 for female; 0 for male);
- Age ($X_{2-5}$), specified with 5 classes, through 4 dummies (baseline "Less than 40");
- Academic rank ($X_{6-7}$), specified by 2 dummies (baseline "Assistant professors");
- Total publications authored by the professor in the period under observation ($X_8$);
- Level of specialization of the scientific activity of the professor, specified by 2 dummies:
    - ✓ "Highly diversified" ($X_9$), 1 if the papers falling in the prevalent subject category of the professor[3] is less than 40% of total publications; 0, otherwise;
    - ✓ "Highly specialized" ($X_{10}$), 1 if the papers falling in the prevalent subject category of the professor is more than 75% of total publications; 0, otherwise;[4]

---

[1] The harmonic average of precision and recall (F-measure) of authorships, as disambiguated by the algorithm, is around 97% (2% margin of error, 98% confidence interval).
[2] Identified through the manual scrutiny and unification of all bibliographic addresses with "Italy" as affiliation country.
[3] For a thorough explanation of the bibliometric approach to discriminate specialized from diversified research, we refer the reader to Abramo, D'Angelo, and Di Costa (2017).
[4] The chosen thresholds (40% for $X_8$, and 75% for $X_9$) allow for equal partitions of the dataset (one third of of highly diversified professors, and one third of highly specialized ones).

Contextual covariates
- Environment-Peers behaviour ($X_{11}$), specified by a dummy variable (1 in case of a colleague in the same university and SDS of the professor, co-authoring publications with industry; 0, otherwise);
- Institutional control ($X_{12}$), specified by a dummy variable (0, for public universities; 1, for private ones);
- University scope ($X_{13}$), specified by a dummy variable (1, for "Polytechnics" and "Special Schools for Advanced Studies, SS"; 0, otherwise);[5]
- University size in the ADU of the professor ($X_{14-15}$), specified with 3 classes through 2 dummies ("Big", for universities with a research staff in the UDA of the professor, above 80 percentile in the national ranking; "Medium", with a research staff between 50 and 80 percentile);
- University location ($X_{16-19}$), specified with 5 geographical macro-areas, by 4 dummies (baseline "Islands").

In order to control for the research discipline effects, we also consider other 9 dummies related to the ten UDAs under observation.

The logit regression was applied separately to the two period under observation. In detail, values of the variables referred to the first period are measured at 31/12/2009, while the ones of the second period are measured at 31/12/2013.

**Results**

In the first period, professors with at least one publication co-authored with industry represented 17.2 percent of the 2010-2013 dataset (Table 1). In the second period, their share raised to 25.5 percent. Among the 27,323 professors on staff in both periods, 10.0 percent collaborated with industry in both periods; 8.1 percent collaborated in the first period only, and 16.0 percent collaborated in the second period only. Overall, the share of professors on staff in both periods, and who collaborated with industry in the first period was 18.1 percent; in the second period, 26.0 percent.

The other two subsets considered in Table 1 (A not B, B not A), consist respectively of professors who left the academia in the second period, and who joined it in the second period. Among the former, the share of those who co-authored at least one publication with industry in 2010-2013 was 11.3 percent. Among the latter, the share almost doubled (21.6%).

We can say then, that i) the share of professors on staff in both periods, collaborating with industry increased by 8.0 percent; and ii) the new entries (in the second period) showed a propensity to collaborate with industry which almost doubled that of professors approaching retirement.[6]

---

[5] The Italian Minister of University and Research (MUR) recognizes a total of 96 universities as having the authority to issue legally recognized degrees. Of these, 29 are small, private, special-focus universities, of which 13 offer only e-learning, 67 are public and generally multi-disciplinary universities. Three of them are Polytechnics and six are *Scuole Superiori* (Special Schools for Advanced Studies), devoted to highly talented students, with very small faculties and tightly limited enrolment per degree program. In the overall system, 94.9% of faculty are employed in public universities (0.5% in *Scuole Superiori*).

[6] Almost all professors present only in the first period, were not there in the second period because of age limits. In Italy, it is quite rare that a professor quits academia for other reasons (Abramo, D'Angelo, & Rosati, 2016).

*Table 1. Breakdown of professors in the dataset*

| Set* | No. | Category | Share |
|---|---|---|---|
| A | 31643 | Collaborating | 17.2% |
| B | 31392 | Collaborating | 25.5% |
| A ∩ B | 27323 | Collaborating in both periods | 10.0% |
| | | Collaborating only in the second period | 16.0% |
| | | Collaborating only in the first period | 8.1% |
| | | Never collaborating | 65.8% |
| A not B | 4320 | Collaborating with industry | 11.3% |
| B not A | 4069 | Collaborating with industry | 21.6% |

\* "A" = professors in the 2010-2013 dataset; "B" = professors in the 2014-2017 dataset

Table 2 reports the typical profile of university professors collaborating with private companies, identified by the concentration index on each individual and contextual trait.[7] The left side refers to the profile emerging from the data of the first four-year period; the right side, to the second one.

2010-2013, the typical academic active in research collaboration with industry is male, 40-45 years old, a full professor, with highly diversified research activity. This professor operates within a group of peers who likewise collaborate with industry, and belongs to a medium-sized public university, typically a polytechnic or SSs located in northwestern Italy.

In 2014-2017, the profile is very similar. The only significant differences are in the age of the academic and size of their home university. As for the age, the prevailing trait is below 40 years old: in other words, in the second period, interaction with industry extends to and features academics younger than other aged colleagues. As for the size, the shift is from medium to big, but in both period the association is not statistically significant. Other traits remain unchanged, although variation in the absolute value of the concentration index reveals a weakening/strengthening of the characteristic trait between the two periods.

Since this purely descriptive analysis does not take into account the simultaneous effects of all covariates on the independent variable, in the following we conduct an inferential analysis using a logit regression model, as illustrated in the previous section. Table 3 shows the average values of the model variables, at overall level. Some significant differences between the two periods, for individual covariates, deserve an underlining:
- Average age of professors increased, in particular the incidence of the elder courts (for example, the over sixties' share raised from 12.6 percent to 14.3 percent of total);
- Average number of publications per professors increased (from 14.60 to 19.37);
- The propensity to diversify research activity increased. Highly diversified academics raised from 14.3 percent to 19.1 percent. Specularly, the share of specialized researchers decreased from 31.3 percent to 25.2 percent).

We remind the reader that the two datasets partly overlap, with about 87 percent of professors belonging to both. As a consequence, variations between the two periods can be explained partly by a behavioural change of incumbent professors, and partly by different attitudes and interests of younger newly hired professors as compared to their older peers who retired. In fact, in the second period, the share of associate professors, and more noticeably of full professors, dropped in favour of that of assistant professors. In spite of that, the average age of faculty increased, in particular that of the elder courts (the share of over sixty years old professors raised from 12.6 percent to 14.3 percent). Thanks to the new entries and exits, gender unbalance improved, as the share of female professors raised from 32.3 percent to 33.9 percent.

---

[7] The concentration index is the ratio of two ratios. Example: for the group variable "gender", the prevailing trait "male" shows a concentration index of 1.058 for 2010-2013 data, since males compose 71.66% of total researchers co-authoring publications with industry, and 67.75% of the total population, therefore 71.66/67.75=1.058.

In fact, in the subset "B not A" of Table 1, the share of females is 37.6 percent, while in the subset "A not B", it is 25.3 percent.

*Table 2: Profiling of the Italian academic co-authoring publications with industry*

| Group variable | 2010-2013 | | | 2014-2017 | | |
|---|---|---|---|---|---|---|
| | Prevailing trait | Concentr. index | Pearson chi$^2$ | Prevailing trait | Concentr. index | Pearson chi$^2$ |
| Gender | Male | 1.058 | 46.1*** | Male | 1.076 | 120.0*** |
| Age | 40-45 years old | 1.073 | 54.9*** | Below 40 years old | 1.132 | 132.1*** |
| Academic rank | Full professor | 1.160 | 78.2*** | Full professor | 1.150 | 80.9*** |
| UDA | 9 - Ind.+Infor. Engineering | 1.661 | 1.0e+03*** | 9 - Ind.+Infor. Engineering | 1.777 | 1.9e+03*** |
| Scientific activity | Highly diversified | 1.328 | 356.5*** | Highly diversified | 1.220 | 306.3*** |
| Environment | Peers collaborating | 1.432 | 1.6e+03*** | Peers collaborating | 1.265 | 1.5e+03*** |
| University type | Public | 1.005 | 4.6** | Public | 1.012 | 35.6*** |
| | Polytechnic or SS | 1.619 | 152.3*** | Polytechnic or SS | 1.676 | 309.0*** |
| University size | Medium | 1.014 | 0.7 | Big | 1.005 | 1.9 |
| University location | North-west | 1.262 | 167.2*** | North-west | 1.148 | 149.1*** |

*Statistical significance: \*p-value <0.10, \*\*p-value <0.05, \*\*\*p-value <0.01*

*Table 3: Average values of the regression model variables*

| Variable group | | | 2010-2013 | 2014-2017 |
|---|---|---|---|---|
| Response | Y | Co-authorships with industry | 0.172 | 0.255 |
| Gender | $X_1$ | Female | 0.323 | 0.339 |
| Age | $X_2$ | 40-45 | 0.200 | 0.190 |
| | $X_3$ | 46-52 | 0.244 | 0.252 |
| | $X_4$ | 53-60 | 0.229 | 0.258 |
| | $X_5$ | Over 60 | 0.126 | 0.143 |
| Academic rank | $X_6$ | Associate prof. | 0.284 | 0.281 |
| | $X_7$ | Full prof. | 0.247 | 0.217 |
| Producivity | $X_8$ | Tot. publications | 14.60 | 18.37 |
| Scientific activity | $X_9$ | Highly diversified | 0.143 | 0.191 |
| | $X_{10}$ | Highly specialized | 0.313 | 0.252 |
| Environment | $X_{11}$ | Peers behaviour | 0.561 | 0.660 |
| University type | $X_{12}$ | Private | 0.034 | 0.039 |
| | $X_{13}$ | Polytechnic or SS | 0.057 | 0.059 |
| University size | $X_{14}$ | Medium | 0.329 | 0.321 |
| | $X_{15}$ | Large | 0.598 | 0.610 |
| University location | $X_{16}$ | South | 0.200 | 0.200 |
| | $X_{17}$ | Center | 0.254 | 0.250 |
| | $X_{18}$ | Northeast | 0.197 | 0.196 |
| | $X_{19}$ | Northwest | 0.239 | 0.244 |

Findings of the logit regressions applied in both periods to investigate the drivers of academic engagement in public-private research collaboration, are reported in Table 4.

The model estimation appears satisfactory. For the 2014-2017 model, the mean VIF is 2.49, with maximum (7.72) for the covariate "University size – Big" ($X_{15}$), which excludes the presence of multicollinearity that could disturb the estimation of the coefficients. The value of under ROC area (AUC) is 0.737, which indicates good ability of the model to correctly classify professors, discriminating the propensity to collaborate with companies[8] in function of individual traits and context.

The estimated coefficients of the regression model are expressed in terms of odds ratios: the reference value is equal to one and indicates that the independent variable considered has no effect on the dependent variable, i.e. on the probability that a professor has or has not collaborated with private companies. For values above one, the variable instead has a positive marginal effect, and vice versa.

Data in Table 4 indicate that all the covariates but very few have a statistically significant effect. In particular, "University scope" ($X_{13}$) does not show a significant marginal effect on academics propensity to collaborate with industry in either periods.

Confirming previous literature (Weerasinghe & Dedunu, 2020; Tartari & Salter, 2015; Calvo, Fernández-López, & Rodeiro-Pazos, 2019), among the individual characteristics, gender has a non-marginal effect: women show a lower propensity to collaborate with private companies, decreasing in the second period (from -10.2 percent to -13.0 percent). Age as well shows a systematically negative impact on the response variable of the proposed model, invariable over time, with all odds ratios below 1 and decreasing in a similar way for the two periods.

As for the academic rank, it shows a positive effect on the propensity for collaborating with industry, slightly decreasing in the second period: full and associate professors show a higher propensity to collaborate as compared to assistant professors, respectively by 86 percent and 39 percent in 2010-2013, vs 77 percent and 33 percent in 2014-2017.

Regarding the level of research diversification/specialization, the analysis reveals a marginal effect increasing over time: in the first period, a highly diversified scientific profile had 32 percent higher probability of collaborating with companies than an "intermediate" profile; the effect raised to +37% in the second period. Specularly, the dual covariant (highly specialized profile) has a negative impact on the propensity to collaborate.

Finally, the effect of what can be considered an exposure variable (the number of total publications), although statistically significant, is very limited in both periods (odds ratio 1.012 in 2010-2013, and 1.010 in 2014-2017).

Among all covariates under examination, "Peers collaborating" ($X_{11}$) presents the highest odds ratio. Confirming the indication of Tartari, Perkmann, and Salter (2014) the presence of colleagues from the same field, actively collaborating with companies, is the most important driver for an academic, even though in the second period its marginal effect substantially decreases (odds ratio from 3.472 in 2010-2013 to 2.834 in 2014-2017). Among contextual covariates, the trait "public" of a university has a positive effect on the propensity of faculty to collaborate, probably because financial resources for research are lower than in private universities. Furthermore, all others being equal, in large-sized university the faculty propensity to collaborate is in the first period 33.2 percent (-20.5 percent in the second period) lower than in small-sized ones. As for geographic localization, in the second period, the coefficients of the four dummies considered indicate an increase in the propensity to collaborate with latitude (i.e. towards northern Italy, where the concentration of private R&D is the highest); in the first period, only the dummy North_West presents a statistically significant effect. It seems that along time findings meet expectations, that is for university professors, interaction with companies depends on the level of concentration of industrial activities in general and, in particular, of knowledge-intensive industry.

---

[8] The AUC analysis evaluates a classifier's ability to discern between true positives and false positives. In our case, the AUC value, between 0 and 1, is equivalent to the probability that the result of the logit classifier applied to a researcher randomly extracted from the group of those who collaborated with industry is higher than that obtained by applying it to a researcher randomly extracted from the group of those who did not collaborate (Bowyer, Kranenburg, & Dougherty, 2001).

*Table 4: The main drivers of the propensity to collaborate with industry by Italian professors. Logit regression, dependent variable: 1 in case of publications in co-authorship with industry, 0, otherwise*

| Variable group | | | 2010-2013 | 2014-2017 |
|---|---|---|---|---|
| | | Const. | 0.040*** | 0.061*** |
| Gender | $X_1$ | Female | 0.898*** | 0.870*** |
| Age | $X_2$ | 40-45 | 0.916* | 0.960 |
| | $X_3$ | 46-52 | 0.787*** | 0.768*** |
| | $X_4$ | 53-60 | 0.617*** | 0.626*** |
| | $X_5$ | Over 60 | 0.449*** | 0.470*** |
| Academic rank | $X_6$ | Associate prof. | 1.392*** | 1.332*** |
| | $X_7$ | Full prof. | 1.862*** | 1.775*** |
| Producivity | $X_8$ | Total publications | 1.012*** | 1.010*** |
| Scientific activity | $X_9$ | Highly diversified | 1.320*** | 1.369*** |
| | $X_{10}$ | Highly specialized | 0.611*** | 0.631*** |
| Environment | $X_{11}$ | Peers collaborating | 3.472*** | 2.834*** |
| University type | $X_{12}$ | Private | 0.841* | 0.825** |
| | $X_{13}$ | Polytechnic or SS | 0.918 | 1.062 |
| University size | $X_{14}$ | Medium | 0.831*** | 0.892* |
| | $X_{15}$ | Big | 0.668*** | 0.795*** |
| University location | $X_{16}$ | South | 0.905 | 1.208*** |
| | $X_{17}$ | Center | 1.008 | 1.352*** |
| | $X_{18}$ | North_East | 1.094 | 1.397*** |
| | $X_{19}$ | North_West | 1.337*** | 1.464*** |
| | | Number of obs | 31643 | 31392 |
| | | LR chi2(28) | 3394.67 | 4189.84 |
| | | Prob > chi2 | 0.000 | 0.000 |
| | | Log likelihood | -12589.9 | -15661.6 |
| | | Pseudo $R^2$ | 0.119 | 0.118 |

*Statistical significance: \*p-value <0.10, \*\*p-value <0.05, \*\*\*p-value <0.01*

**Conclusions**

In the so called knowledge-based economy we live in, knowledge is increasingly becoming a distinctive competence to achieve the competitive advantage of companies, and to sustain the economic growth of nations. Universities, as the loci of new knowledge creation, have undergone in the last decades an increasing pressure to devote a special care to the technology transfer function (Todorovic, McNaughton, & Guild, 2011; Etzkowitz & Leydesdorff, 1995; Etzkowitz, 1983), which has become their third mission, alongside the traditional ones of education and research (Etzkowitz & Leydesdorff, 1995). When we refer to public-private technology transfer we generally think the exploitation by industry of results of research conducted at universities and public research organizations. A quicker and more effective way to pursue the same results is the co-creation of new knowledge through joint public-private research projects. To that aim, a number of incentive schemes have been put into place (Davenport, Davies, & Grimes, 1998; Debackere & Veugelers, 2005).

Understanding the drivers underlying the academics' propensity to engage in collaboration with industry is critical in formulating and targeting such incentive systems to maximize their effectiveness. Also important is to be aware of how the relative weight of such drivers vary over time, especially when setting performance indicators and reward systems, or when assessing the effectiveness of the policy initiatives.

Incentives need to overcome disincentives that are at play, first of all transactions costs (Belkhodja & Landry, 2005; Drejer & Jørgensen, 2005), which grow with the cultural and cognitive distance of members of the university-industry research team (Abramo, D'Angelo, Di Costa, & Solazzi, 2011). Furthermore, it has been demonstrated that on average public-private co-authored publications have lower impact than intra-university co-authored publications, and represent a lower share of highly-cited publications (Abramo, D'Angelo, & Di Costa, 2020). Because academics are increasingly subject to evaluation of their scientific activity, this awareness might constitute an additional deterrent.

In this study, we show that in the Italian academia, the share of professors collaborating with industry raised from 17.2 percent in 2010-2013 to 25.5 percent in 2014-2017. This increase is explained partly by a higher propensity to collaborate with industry by incumbents, and partly by the higher propensity of new hired professors as compared to their older colleagues who retired. This evidence can be partly due to the increasing emphasis on the importance of the university's so-called third mission. Researchers are experiencing increasing pressure to open up their research agendas to the needs of the local and national production system. There are no yet specific incentives put in place by the policy maker in this sense, but in the current national evaluation exercise, the entrepreneurial and technology transfer activities of universities are going to be assessed for the first time, in view of their possible use in the performance-based research funding scheme adopted by the Ministry of Universities and Research. At the same time, it is also possible that the increasing recourse to collaboration with private organizations has been caused by the progressive thinning of the resources made available to public research institutions, because of the budget constraints due to the tough economic situation in Italy.

The typical profile of the academic collaborating with industry is unchanged: it is a male, full professors, relatively young, with highly diversified research activity, operating within a disciplinary team made of colleagues with a high propensity themselves to collaborate with industry, and on staff in a public university located in northwestern Italy.

The relative importance of the drivers of academics' engagement in research collaboration with industry remains quite stable. The variations of the marginal effects of such personal traits as gender and age are quite modest. More noticeable is the increase of the marginal effect of research diversification, and the decrease of the effect of academic rank.

With regard to contextual drivers, the variable showing the highest weight, i.e. the presence of peers with high propensity to collaborate with industry, undergoes a substantial reduction. University size loses weight as well, while localization gains it.

In interpreting results, caution is recommended due to the intrinsic limits of all bibliometric approaches to the analyse of cross-sector collaborations, not all research collaborations lead to an indexed publication, and not all joint co-authored publications reflect a real public-private collaboration.

Follow on research might concern i) the measurement of publications' impact to verify whether there occurs a trade-off between the increase of the frequency of collaborations and the quality of their outputs; ii) a field level analysis to verify to what extent results vary across fields; and iii) the extension of the analysis to other time periods with panel data.

Future research might also inquire into the geographic proximity effect on the intensity of cross-sector research collaborations, as compared with intra-sector collaborations. Finally, the methodology can be easily applied to other nations, whereby natives can easily distinguish and reconcile public and private affiliations. This would allow cross-country comparisons for a better understanding of the phenomenon.

**References**

Abramo, G., D'Angelo, C. A., & Murgia, G. (2013). The collaboration behaviours of scientists in Italy: A field level analysis. *Journal of Informetrics*, *7*(2), 442–454. DOI: 10.1016/j.joi.2013.01.009

Abramo, G., D'Angelo, C.A., & Di Costa, F. (2017). Specialization versus diversification in research activities: the extent, intensity and relatedness of field diversification by individual scientists. *Scientometrics*, 112(3), 1403-1418. DOI: 10.1007/s11192-017-2426-7

Abramo, G., D'Angelo, C.A., & Di Costa, F. (2019). Authorship analysis of specialized vs diversified research output. *Journal of Informetrics,* 13(2), 564-573. DOI: 10.1016/j.joi.2019.03.004

Abramo, G., D'Angelo, C.A., & Di Costa, F. (2020). The relative impact of private research on scientific advancement. Working paper, http://arxiv.org/abs/2012.04908, last accessed on 17 December 2020.

Abramo, G., D'Angelo, C.A., & Rosati, F. (2016). A methodology to measure the effectiveness of academic recruitment and turnover. *Journal of Informetrics*, 10(1), 31-42. DOI: 10.1016/j.joi.2015.10.004

Abramo, G., D'Angelo, C.A., Di Costa, F., & Solazzi, M. (2011). The role of information asymmetry in the market for university-industry research collaboration. *The Journal of Technology Transfer,* 36(1), 84-100. DOI: 10.1007/s10961-009-9131-5.

Balconi, M., & Laboranti, A. (2006). University-industry interactions in applied research: The case of microelectronics. *Research Policy*, 35(10), 1616-1630. DOI: 10.1016/j.respol.2006.09.018

Bayer, A. E., & Smart, J. C. (1991). Career publication patterns and collaborative "styles" in American academic science. *Journal of Higher Education*, 62(6), 613–636.

Bekkers, R., & Bodas Freitas, I. M. (2008). Analysing knowledge transfer channels between universities and industry: To what degree do sectors also matter? *Research Policy,* 37(10), 1837-1853. DOI: 10.1016/j.respol.2008.07.007

Belkhodja, O., & Landry, R. (2005). The Triple Helix collaboration: Why do researchers collaborate with industry and the government? What are the factors influencing the perceived barriers? *5th Triple Helix Conference*, 1–48.

Berbegal-Mirabent, J., Sánchez García, J. L., & Ribeiro-Soriano, D. E. (2015). University-industry partnerships for the provision of R&D services. *Journal of Business Research, 68*(7), 1407-1413. DOI: 10.1016/j.jbusres.2015.01.023

Boschini, A., & Sjögren, A. (2007). Is team formation gender neutral? Evidence from coauthorship patterns. *Journal of Labor Economics*, 25(2), 325–365. DOI: 10.1086/510764

Bowyer, K., Kranenburg, C., & Dougherty, S. (2001). Edge detector evaluation using empirical ROC curves. *Computer Vision and Image Understanding, 84*(1), 77-103. DOI: 10.1006/cviu.2001.0931

Bozeman, B., & Gaughan, M. (2011). How do men and women differ in research collaborations? An analysis of the collaborative motives and strategies of academic researchers. *Research Policy*, 40(10), 1393–1402. DOI: 10.1016/j.respol.2011.07.002

Calvo, N., Fernández-López, S., & Rodeiro-Pazos, D. (2019). Is university-industry collaboration biased by sex criteria? *Knowledge Management Research and Practice, 17*(4), 408-420. DOI: 10.1080/14778238.2018.1557024

D'Angelo, C. A., Giuffrida, C., & Abramo, G. (2011). A heuristic approach to author name disambiguation in large-scale bibliometric databases. *Journal of the American Society for Information Science and Technology*, 62(2), 257-69. DOI: 10.1002/asi.21460.

D'Este, P., & Patel, P. (2007). University-industry linkages in the UK: What are the factors underlying the variety of interactions with industry? *Research Policy*, 36(9), 1295–1313. DOI: 10.1016/j.respol.2007.05.002

Davenport, S., Davies, J., & Grimes, C. (1998). Collaborative research programmes: building trust from difference. *Technovation*, 19(1), 31–40. DOI: 10.1016/S0166-4972(98)00083-

2.
Debackere, K., & Veugelers, R. (2005). The role of academic technology transfer organizations in improving industry science links. *Research Policy*, *34*(3), 321–342. https://doi.org/10.1016/j.respol.2004.12.003

Di Gregorio, D., Shane, S. (2003). Why do some universities generate more start-ups than others? *Research Policy*, 32(2), 209–227. DOI: 10.1016/S0048-7333(02)00097-5

Etzkowitz, H. (1983). Entrepreneurial scientists and entrepreneurial universities in American academic science. *Minerva*, *21*(2–3), 198–233. https://doi.org/10.1007/BF01097964.

Etzkowitz, H., & Leydesdorff, L. (1995). The Triple Helix: University - Industry - Government Relations A Laboratory for Knowledge Based Economic Development. *EASST Review*, *14*, 14–19.

Fan, X., Yang, X., & Chen, L. (2015). Diversified resources and academic influence: patterns of university–industry collaboration in Chinese research-oriented universities. *Scientometrics*, *104*(2), 489–509. DOI: 10.1007/s11192-015-1618-2

Garcia, R., Araújo, V., Mascarini, S., Santos, E. G., & Costa, A. R. (2020). How long-term university-industry collaboration shapes the academic productivity of research groups. *Innovation: Organization and Management, 22*(1), 56-70. DOI: 10.1080/14479338.2019.1632711

Giuri, P., Munari, F., Scandura, A., & Toschi, L. (2019). The strategic orientation of universities in knowledge transfer activities. *Technological Forecasting and Social Change, 138*, 261-278. DOI: 10.1016/j.techfore.2018.09.030

He, Z., Geng, X., & Campbell-Hunt, C. (2009). Research collaboration and research output: A longitudinal study of 65 biomedical scientists in a New Zealand university. *Research Policy, 38*(2), 306-317. DOI: 10.1016/j.respol.2008.11.011

Iorio, R., Labory, S., & Rentocchini, F. (2017). The importance of pro-social behaviour for the breadth and depth of knowledge transfer activities: An analysis of italian academic scientists. *Research Policy, 46*(2), 497-509. DOI:10.1016/j.respol.2016.12.003

Katz, J. S., & Martin, B. R. (1997). What is research collaboration? *Research Policy*, *26*(1), 1–18. DOI: 10.1016/S0048-7333(96)00917-1

Lee, S., & Bozeman, B. (2005). The impact of research collaboration on scientific productivity. *Social Studies of Science, 35*(5), 673-702. DOI: 10.1177/0306312705052359

Llopis, O., Sánchez-Barrioluengo, M., Olmos-Peñuela, J., & Castro-Martínez, E. (2018). Scientists' engagement in knowledge transfer and exchange: Individual factors, variety of mechanisms and users. *Science and Public Policy, 45*(6), 790-803. DOI: 10.1093/scipol/scy020

Mansfield, E. (1995). Academic research underlying industrial innovations: sources, characteristics, and financing. *Review of Economics and Statistics*, 77(1), 55-65

Mansfield, E., & Lee, J.Y. (1996). The modern university: contributor to industrial innovation and recipient of industrial R&D support. *Research Policy*, 25 (7), 1047-1058. DOI: 10.1016/S0048-7333(96)00893-1

Perkmann, M., Tartari, V., McKelvey, M., Autio, E., Broström, A., D'Este, P., . . . Sobrero, M. (2013). Academic engagement and commercialisation: A review of the literature on university-industry relations. *Research Policy, 42*(2), 423-442. DOI: 10.1016/j.respol.2012.09.007

Schartinger, D., Schibany, A., & Gassler, H. (2001). Interactive relations between university and firms: empirical evidence for Austria. *Journal of Technology Transfer*, 26, 255–268.

Shane, S. (2004). Encouraging university entrepreneurship? The effect of the Bayh-Dole Act on university patenting in the United States. *Journal of Business Venturing*, *19*(1), 127–151. DOI: 10.1016/S0883-9026(02)00114-3

Tartari, V., & Salter, A. (2015). The engagement gap: Exploring gender differences in university - industry collaboration activities. *Research Policy, 44*(6), 1176-1191. DOI: 10.1016/j.respol.2015.01.014

Tartari, V., Perkmann, M., & Salter, A. (2014). In good company: The influence of peers on industry engagement by academic scientists. *Research Policy, 43*(7), 1189-1203. DOI:10.1016/j.respol.2014.02.003

Thune, T., Reymert, I., Gulbrandsen, M., & Aamodt, P. O. (2016). Universities and external engagement activities: Particular profiles for particular universities? *Science and Public Policy, 43*(6), 774-786. DOI: 10.1093/scipol/scw01.

Todorovic, Z. W., McNaughton, R. B., & Guild, P. (2011). ENTRE-U: An entrepreneurial orientation scale for universities. *Technovation*, *31*(2–3), 128–137. https://doi.org/10.1016/j.technovation.2010.10.009

Traoré, N., & Landry, R. (1997). On the determinants of scientists' collaboration. *Science Communication*, *19*(2), 124–140. DOI: 10.1177/1075547097019002002

Ubfal, D., & Maffioli, A. (2011). The impact of funding on research collaboration: Evidence from a developing country. *Research Policy*, *40*(9), 1269–1279. DOI: 10.1016/j.respol.2011.05.023

Weerasinghe, I. M. S., & Dedunu, H. H. (2020). Contribution of academics to university–industry knowledge exchange: A study of open *innovation* in Sri Lankan universities. *Industry and Higher Education,* DOI: 10.1177/0950422220964363

Zhao, Z., Broström, A., & Cai, J. (2020). Promoting academic engagement: University context and individual characteristics. *Journal of Technology Transfer, 45*(1), 304-337. DOI: 10.1007/s10961-018-9680-6